# All About Phishing
# Exploring User Research through a Systematic Literature Review

S. Das, A. Kim, Z. Tingle, and C. Nippert-Eng

School of Informatics, Computing, and Engineering

Indiana University Bloomington

E-mail: {sancdas, anykim, zatingle, cnippert}@iu.edu

## Abstract

Phishing is a well-known cybersecurity attack that has rapidly increased in recent years. It poses legitimate risks to businesses, government agencies, and all users due to sensitive data breaches, subsequent financial and productivity losses, and social and personal inconvenience. Often, these attacks use social engineering techniques to deceive end-users, indicating the importance of user-focused studies to help prevent future attacks. We provide a detailed overview of phishing research that has focused on users by conducting a systematic literature review of peer-reviewed academic papers published in ACM Digital Library. Although published work on phishing appears in this data set as early as 2004, we found that of the total number of papers on phishing (N = 367) only 13.9% ($n = 51$) focus on users by employing user study methodologies such as interviews, surveys, and in-lab studies. Even within this small subset of papers, we note a striking lack of attention to reporting important information about methods and participants (e.g., the number and nature of participants), along with crucial recruitment biases in some of the research.

## 1. Introduction

Phishing is one of the most effective and well-known cyber threats, leading to millions of compromised credentials and contributing to 90% of data breaches (Retruster.com, 2019). Phishing scams are becoming increasingly more deceptive with sophisticated attacks that are able to manipulate end-users through, for instance, spoofed websites, targeted emails, and fake phone calls. Highly publicized phishing scam incidents, like the *John Podesta Case*, show that, despite a user's technical knowledge and background, anyone can fall victim to these attacks, and the consequences may be profound (Uchill, 2016).

Kay et al. report that the term "Phishing" originated in 1996, when hackers were stealing online data from American accounts. These hackers used emails as "hooks" to catch their "fish" from the "sea" of internet users (Kay, 2004). Today, there are a number of known types of phishing attacks, such as Deceptive Phishing, Malware Based Phishing, Keyloggers and Screenloggers, Session Hacking, Web Trojans, Spear Phishing, Search Engine Phishing, Content Injection Phishing, DNS-Based Phishing, and Vishing (Elledge, 2004). Often, security tools and warnings are provided as a solution to mitigate such attacks (Das et al., 2017). The end-users and their role in this process, however, are surprisingly understudied.

In preparation for our own research to better understand the user and their role in fending off or falling victim to phishing attacks, we conducted an in-depth, systematic literature review of phishing research. In this paper, we report on our findings from this literature review, exploring whether and how human factors appear in this body of work. We provide the methodology and analysis of our research by outlining the protocol we utilized to gather and analyze our data set in sections 3 and 4. We provide a discussion of our findings in section 5. In section 6, we conclude by focusing on future research and the courses of action that are still needed to better understand the user and their motivations and behaviors as they respond to phishing efforts, including a call for researchers to better report on the user component of any future work.

## 2. Background Literature

During a phishing attack, attackers use digital deception to get their victims to reveal confidential information about themselves. The success of the deception depends on how well the attackers mimic legitimate services and contacts and how well the user can distinguish between and/or act appropriately toward what is fake and what is authentic. For instance, there might be a phishing

email with obvious spelling mistakes that signals its illegitimacy. If these spelling mistakes go unnoticed before the recipient clicks, however, the attack will be successful.

The basic and most common form of phishing-related deception is *Identity Deception* (Bakhsh et al., 2008). Here, the deceiver may target a victim and individualize the attack by gathering enough information about the victim to "socially engineer", or personalize, the attack and increase the chances of the user falling victim to it. Another technique, however, provides the intended victim with only generalized information, but the criminals pose as a trusted organization and provide directives to the victim with which they are inclined to comply (Colarik and Janczewski, 2007).

Chou et al. argued that humans are the ones who must be studied in order to stop web-based identity theft, including that which is stolen via phishing (Chou et al., 2004). This position was supported by Tsow et al.'s large-scale user study on phishing, for instance, which analyzed the effect of cyber deception on individuals. The authors found that (dis) similarity in graphic design elements can change users' evaluation of the authenticity of a communique and that the nature of the content plays an important role when users analyze the website (Tsow and Jakobsson, 2007). Such insight becomes even more important in light of Karakasiliotis et al.'s, findings that only 36% of their participants could identify legitimate websites, and only 45% of participants could correctly identify illegitimate websites (Karakasiliotis et al., 2006). Such studies not only address the importance of risk awareness among end-users but also indicate the potential importance of effective risk communication as at least a partial solution in helping users fend off deceptive attacks. As Hong et al. noted, "It doesn't matter how many firewalls, encryption software, certificates, or two-factor authentication mechanisms an organization has if the person behind the keyboard falls for a phish" (2012).

The central role played by the user in phishing attacks is precisely why we wish to better understand the current state of user-centered phishing research, including a wide range of methodological approaches and potentially significant attack attributes. More traditional research methods include those where users might, for instance, provide interview feedback on a prototype's usability (Reeder et al., 2018). We are also interested, however, in more passive methods that nonetheless tell us much about what users are and are not paying attention to when faced with a phishing attack. For instance, Alsharnouby et al. used eye tracking techniques to obtain their data about user attention to site authenticity (2015). Their results showed that gaze time on the browser elements provided a positive impact on the ability to detect phishing websites, while the participant's technical expertise had no significant influence on the detection procedure they followed. Indeed, Dhamija et al. found that visual deception can fool even sophisticated users; a good phishing website fooled 90% of their participants (2006). Standard security indicators are not effective enough, they argued, given their analysis of participants' strategies for detecting a phishing website. Such studies help establish the effectiveness of human-centered research on phishing, as well as justify our focus on the current state of user studies in the phishing literature.

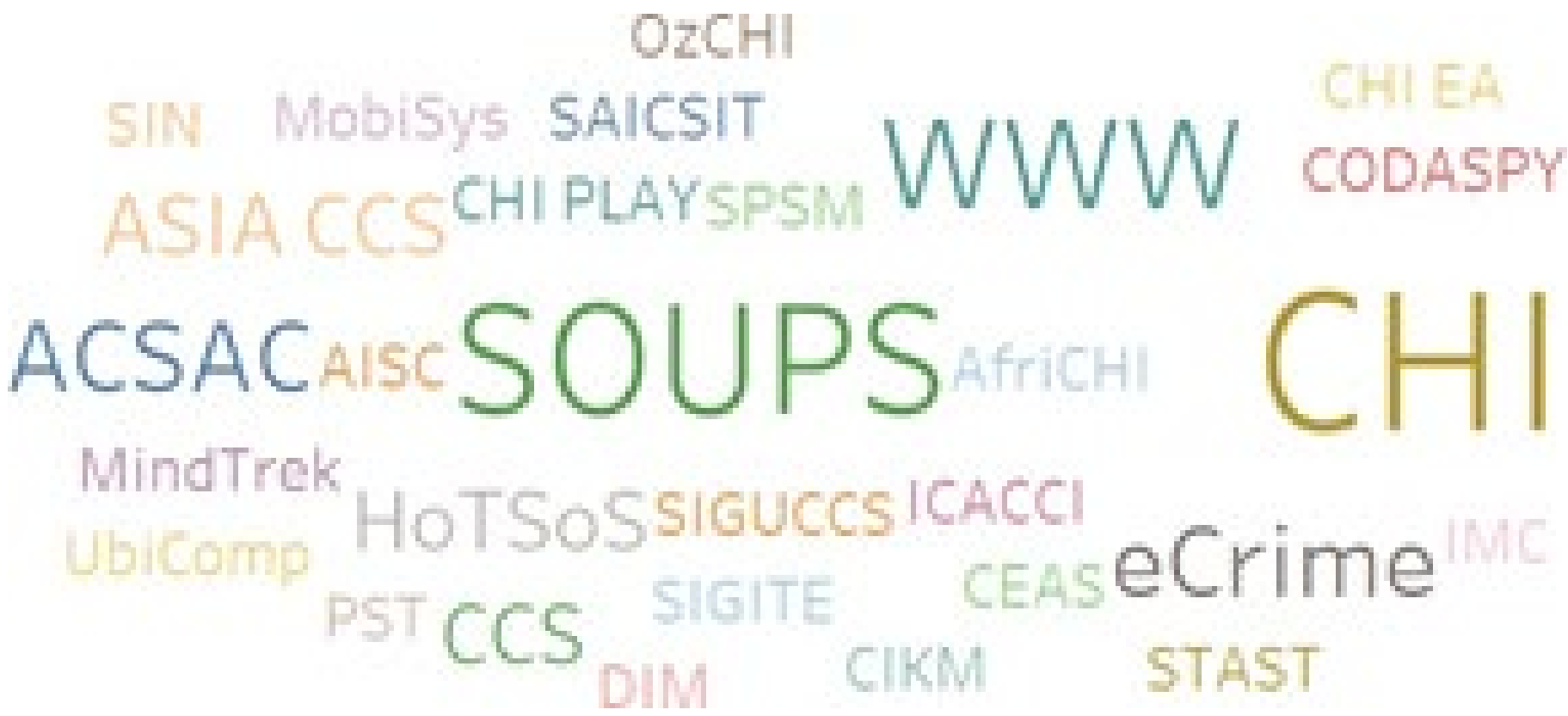

**Figure 1: Word cloud depicting relative representation of conference publication venues in our data set of 51 papers.**

## 3. Study Methodology

Our systematic literature review focused on published research on phishing. We collected our data by starting with all the research publications on phishing that are included in the ACM Digital Library. We performed the data extraction using ACM's export feature and then implemented a qualitative assessment protocol that utilized exclusion and inclusion criteria to generate a set of papers that were appropriate and relevant for our analysis. The result was a total of 367 published papers on phishing, from which we eventually identified 51 papers that included some kind of user study or user evaluation. Three researchers coded the abstract and full text of each of these 51 papers to enable our systematic analysis of the authors' reported research.

## 3.1. Data Collection and Screening

The initial process for our data collection began as a broad search for the term "phishing" in the ACM Digital Library database. This generated 367 papers. From this collected set of papers, we performed abstract and full text screening to identify papers that satisfied our inclusion and exclusion criteria. To be included, a paper needed to be primarily focused on the topic of phishing attacks. Papers were excluded if: (1) they were an extended abstract or a work-in progress, (2) the primary language in which they were written was not English, or (3) they were found not to be related to phishing, even if they mentioned phishing somewhere in the paper. After applying the exclusion criteria on the collected sample of 367 papers, we were left with 253 papers. Three researchers trained in qualitative data coding and analysis independently identified which of these 253 papers included user focused studies. Only these user focused papers made it into our final data set of 51 publications.

Figure 1 shows the word cloud depicting the conference publication venues of our 51 collected papers. Not surprisingly, our dataset is dominated by papers published through the ACM's Computer-Human Interaction (CHI) and Symposium on Usable Privacy and Security (SOUPS) venues. The largest number of papers were from CHI or regional CHI alternatives, such as OzCHI, AfriCHI, CHI Play, and CHI EA (24 papers), while eight papers were from SOUPS, and five each were from CCS and WWW, respectively.

## 3.2. Coding and Analysis

| **Overarching themes observed in the 51 papers through thematic coding** | **Specific phishing attempt attributes observed in user studies** |
|---|---|
| Technical Attributes - form and content-related aspects of the phishing attempts sent to the victims (e.g., emails, fake phone calls) | E.g., Content, Advertisement, Spelling, Readability, Language, URL, Sender's Address, Personalized Contents, Interface, Security Indicators, Graphics, etc. |
| Individual Attributes - personal factors that contributed to users recognizing or falling victim to an attack | E.g., Curiosity, Tiredness, Age, Security Fatigue, Time of Day, Technical Background, Past Experience, Gender, Training, Authority, Relationship with Sender, Social Proof, Risk Perception, etc. |
| Benefits - fake enticements offered in exchange for complying with attackers' desires | E.g., Financial, Job Prospect, Opportunities, Invitation to Event, etc. |
| Threats - fake undesirable consequences used to pressure recipients into complying with attackers' wishes | E.g., By government, Visa-related, Account Closure, Data Leak, Social Failure, Ransom, etc. |

**Table 1: Examples of emergent themes used for qualitative coding of 51 user-focused studies**

Researchers performed thematic coding on the user and technical attributes reported for each study. The coding strategies helped us understand the researchers' focus of attention in these user-focused studies. Table 1 includes a summary of the types of study elements we found. These include a focus on the *technical attributes* of the phishing attempt (largely form- and content-related), the *individual attributes* of the users who were targeted for the phishing attempt (from expertise and frame of mind to demographics), and a focus on the kinds of *benefits* (e.g., "free virus protection software") or *threats* (e.g., "Your password must be updated or you will lose access to your account") used by the attackers to "reel in" the intended phishing victims. In addition to these variables, we also coded study information such as participant age distribution, type of research design, proposed technical and user-focused solutions, and apparent recruitment biases (if any).

Three independent researchers who were trained in qualitative coding generated a set of open codes from a randomly sampled subset of twenty from the collected dataset. These open codes then went through axial and thematic coding, respectively, to generate the themes and attributes mentioned in Table 1. Each paper was then coded independently by at least two researchers. A third researcher reviewed any situation in which the two coders were not in agreement and made the final decision about how the attribute would be coded. The inter-coder reliability (ICR) score after the first iteration of coding and discussion was 23.7%; thus, we went through a second round of discussion, increasing the ICR score to 56.7%. After the final round of discussion, the researchers coded all of the 51 papers, where the ICR score was observed as 87.9%.

## 4. Findings

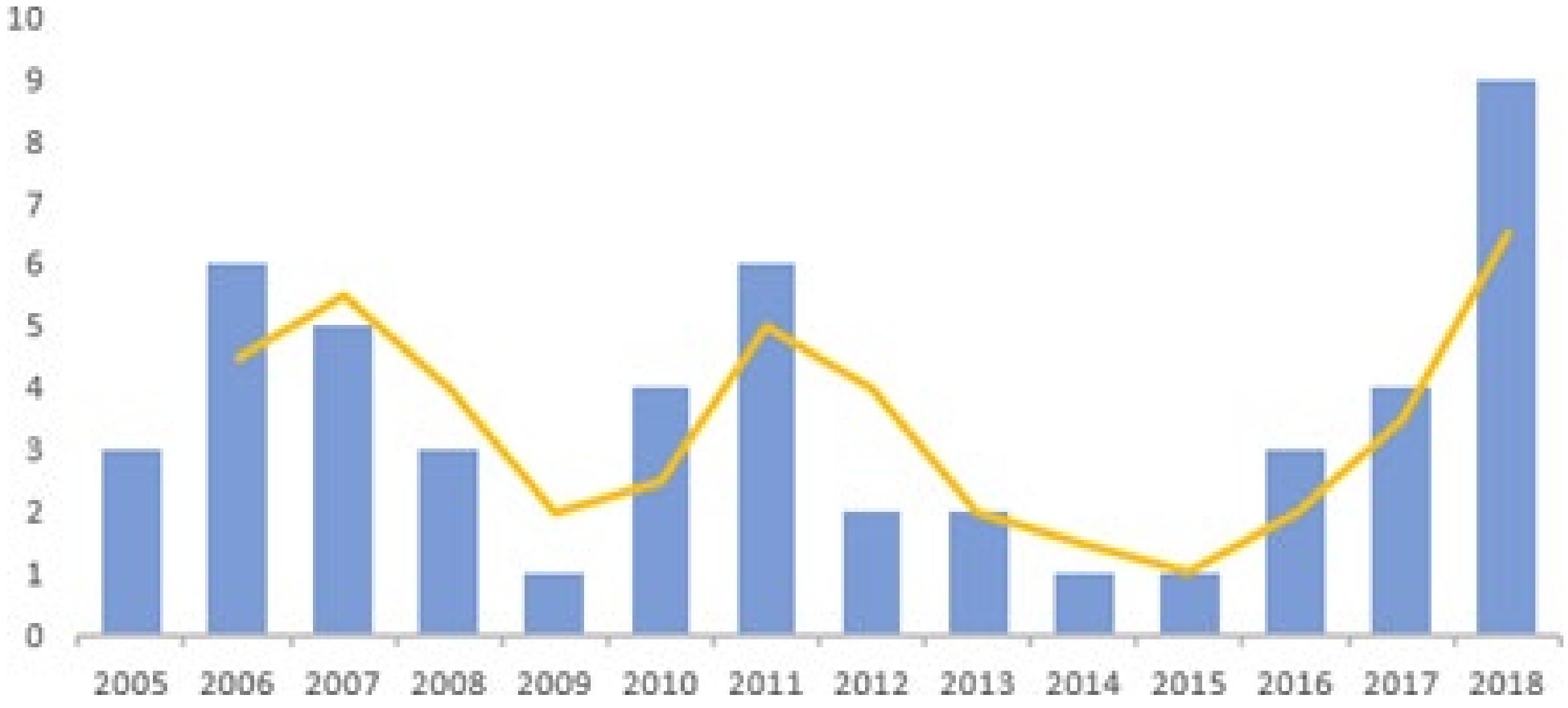

**Figure 2: Number of publications in our data set (n = 51) by year of publication**

Throughout the course of our systematic literature review, we found specific trends in phishing research. Although the term “phishing" was coined in 1996 (Kay, 2004), academic researchers did not begin publishing about phishing until 2004 (Dunham, 2004). The first user-centered study we see in our data set was from 2005 (Garfinkel and Miller, 2005). Figure 2 shows the distribution of the ACM Digital Library’s user focused studies in phishing over the past 15 years. Beginning in 2005 with three papers published in this domain, we see a positive trend that ends in 2018, with eight papers published in the area.

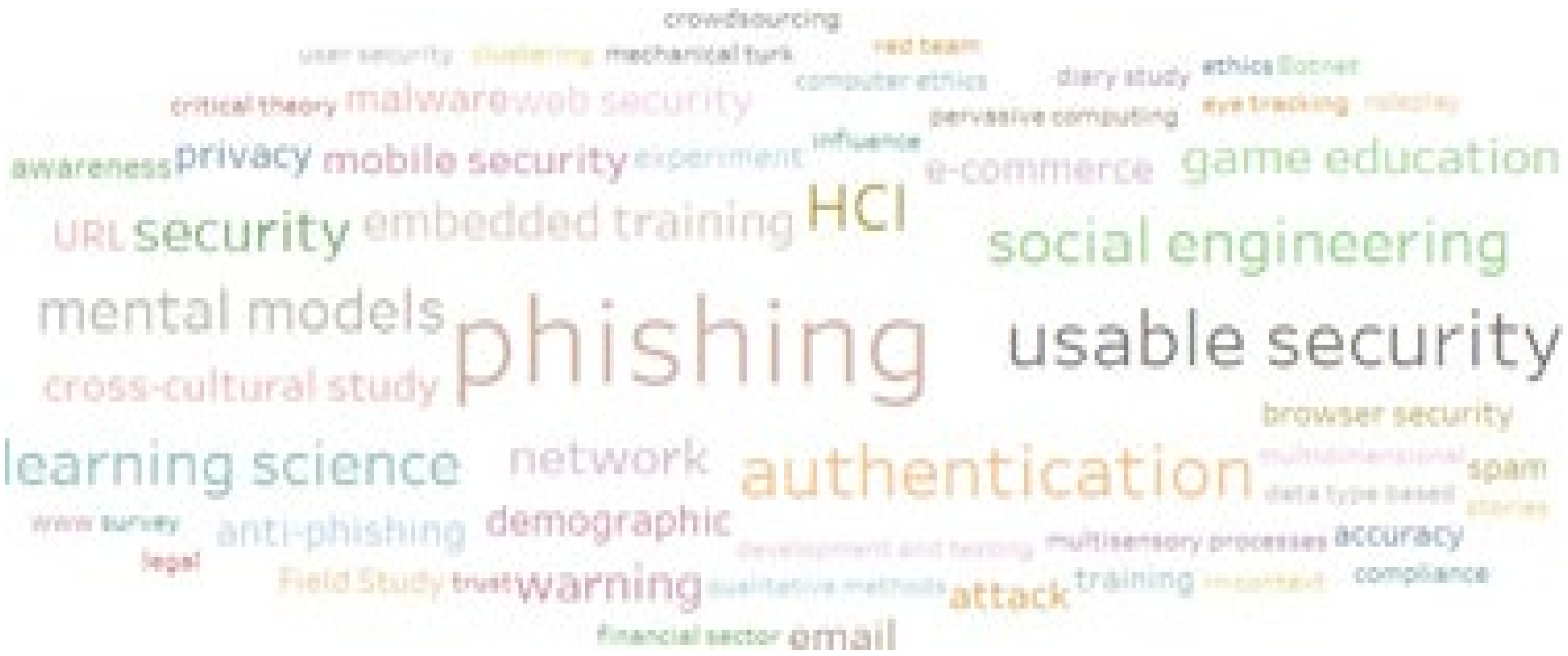

**Figure 3: Word cloud depicting the list of author-selected keywords for our collected data set of 51 papers**

Figure 3 is a word cloud of the authors’ selected keywords from our dataset of 51 user-focused papers. Unsurprisingly, the most common keyword used was “phishing,” included on 34 papers. “Usable security” was the second most frequent (included on eight papers), and “authentication” was the third most common (included on six papers). In the following sections, we discuss the main subjects that received the authors’ attention.

## 4.1. Technical Attributes of Phishing Attacks

Forty out of the 51 collected papers included research that focused on the technical attributes of phishing attacks. We have clustered these attributes into four groups, ranging from the appearance of the phishing attempt (e.g., Tembe et al., 2014; Malisa et al., 2017), such as the look of fake websites and graphical similarities with authentic websites, to the content of the communique (Blythe et al., 2011), including elements such as spelling and official-looking messages sent from a workplace or educational institutes, and to social engineering factors (Meyers et al., 2018) such as the use of personalized data in the phishing attempt, and warnings/indicators of deception (Neupane et al., 2015), such as security tools indicating the authenticity of websites. Attention to the appearance of the phishing communique was the most common research focus throughout the papers. For example, Dhamija et al. conducted a user study that revealed that, despite user intentions to discover a phishing website, many users were not able to look at a website and distinguish legitimate websites from phishing websites (2006). We found that 26 papers focused their research on some combination of attributes that fell into all four categories, while 21 of these papers focused primarily on the appearance of the fake websites/emails.

## 4.2. Individual Attributes of Phishing Attacks

Thirty-seven papers in our data set focus on what we call "individual attributes," which encompass a wide range of the users' personal attributes used by perpetrators to try to influence their intended victims. These include, for example, age, gender, and previous experiences. The goal of a cyber-deception artist is to take advantage of individual human qualities like these and to exploit any personal information. In an attempt to mitigate these risks, training has been consistently identified as an important means of combating successful phishing attempts. Some researchers, for instance, focus on risk communication through videos (Yamanoue et al., 2005; Gupta et al., 2018) or games (Monk et al., 2010; Sheng et al., 2007; Kumaraguru et al., 2010). Wash et al. sought to understand exactly what types of phishing training were most effective in informing users about these attempts (2018). They found that the presenter of the training is key to mitigating phishing attacks, with security professionals being the most effective for training and peers being more effective for telling memorable stories about attempts (Ronda et al., 2008; Downs et al., 2006).

Although rarely studied, these tools can be made more effective by understanding the mental models of the users. For instance, Pandit et al., conducted an extensive user training test with a game called "PHISHY," which was designed based on the mental models of participants. The game resulted in higher rates of identification of phishing links among corporate users, indicating the effectiveness of incorporating the user perspective (Pandit et al., 2018). Phishing research on actors' frames of mind may also aim to better understand that of the attacker. Mehresh et al. proposed a framework to predict an attacker's intent in order to design a stronger and more effective security recovery system (2012). Several other studies, including the one by Almeshekah et al., discuss how deception can be used for security purposes and where planning should be done to understand the attacker's perspective (2016). We believe this is a fruitful area for exploration on both sides of a phishing attack.

## 4.3. Motivations for Participating in Phishing Attacks

Spear phishing is the most common and effective type of phishing, as it focuses on specific individuals, using personal information about its victims. Spear phishing emails may address users by their real name(s) or reference uniquely identifiable information obtained through social engineering techniques (TrendLabs APT Research Team, 2012). Twelve papers in our data set focused on understanding users' risk perceptions of and vulnerability to spear phishing. Spear phishing is successful because attackers manipulate their targets, either by luring users by promising them specific benefits or by coercing users with specific threats (Maurer et al., 2011). These manipulation techniques often lead to impulsive or quick decision making from the end users. One of the most common phishing motivations is the promise of financial benefits to the intended victim. Gao et al. found that many malicious websites attempt to attract users via money or product offers (2010). Attackers often attempt to have users click on their website to earn a free product -- such as an iPhone or video game system -- or to obtain job prospects, such as working online. Surprisingly, however, the details of the content or the reasons why the users were lured into clicking the link were seldom studied by the authors in our data set. Five of the seven papers in this grouping studied this aspect of phishing attacks, exploring how victims are lured in with specific financial benefits (e.g., Hanus et al., 2018) or with career opportunities or invitations to events (Gao et al., 2010).

## 4.4. Study Participants

User studies account for only 13.9% of the ACM Digital Library publications on phishing. Among this small percentage of papers, we found that a number of papers had several reporting issues regarding their participant demographics. Out of the 51 papers, for instance, 15 (about ⅓) of the studies failed to report the total number of participants in their research. A surprising number of papers (37, or a little over ⅔ of the studies) did not mention the age range, gender distribution, or racial/ethnic backgrounds of the participants. Even if the researchers found that none of these participant attributes affected the outcome of the study, these factors should have been reported. Providing basic demographic and technical background information is crucial for understanding the external validity of security focused studies, and for future studies to be able to build on the systematic findings of previous work.

#### 4.4.1. Age Distribution

Only fourteen of the 37 aforementioned papers included any kind of age range of their subjects. Age could be an important factor in usability and phishing studies, as at least some research has shown a significant difference between the way a 20-year-old will perform in a security usability study compared to a 50-year-old (Romano et al., 2013; Chadwick-Dias et al., 2003). We found that ten of these papers studied participants who were somewhere between 39 and 89 years of age, which provided more subject diversity compared to the three studies that only used 18-24 year old participants. Only one study with age diversity actually took advantage of this, however, and examined risk perception differences between older and younger adults (Oliveira et al., 2017).

#### 4.4.2. Gender

Only 27 of the 51 papers provided the gender distribution of their participant population. This is an important factor to report and consider studying further, as research so far suggests that there are differences in how gender affects a user's identification of information from a website (Djamasbi et al., 2007). There may be differences based on gender regarding security behavior and hygiene, in general (Anwar et al., 2017). Gender reporting and diverse gender pools of participants are important for assessing the accuracy and the effectiveness of any security measure.

#### 4.4.3. Race/Ethnicity

Only three of the 51 papers in our data set reported the race/ethnicity of their study participants. None of the papers posited any effect of race/ethnicity on user behavior in either succumbing to or fending off phishing attacks. Previous research has noted an overall lack of racial diversity in security studies (Dev et al., 2018; Das et al., 2019). The paucity of security-related research on participants of diverse racial and ethnic backgrounds is mirrored in our data set.

#### 4.4.4. Methodologies Used

Sixteen of our user-focused studies used a survey to collect data about participant's phishing experience (for example, see Sheng et al., 2010). Interestingly, one of them studied those who provide phishing training through survey data analysis (Wash & Cooper, 2018). Twenty-four papers focused on usability testing of specific tools (for example, see Hart et al., 2011), three of which designed games to educate users about phishing (for example, see Monk et al., 2010). Seven papers performed a controlled phishing attack to understand the risk perception and real-life behavior of users, three of which analyzed participant behavior using interview-based studies (for example, see Kumaraguru et al., 2007). Nine of the collected papers analyzed the content of the phishing attacks, URLs, and spear phishing techniques used by the attackers to phish their victims, while four of these nine papers used surveys to capture user perception of these attacks (for example, see Ndibwile et al., 2018). Two of the studies focused on the theoretical aspect of spear phishing, analyzing the user behavior and personality traits which were targeted by attackers (for example, see Krombholz et al., 2013). Overall, the dearth of interviews and in-depth, qualitative analysis of user behavior is unfortunate. While surveys and usability testing are very important methodological tools to understand user experience and behavior, a combination of interviews and in-depth analysis may yield more beneficial findings.

## 5. Discussion and Implications

Throughout our systematic literature review of user studies in published ACM papers on phishing, we found that there are a number of identifiable trends. The breadth of the research (40 out of 51 papers) concentrates primarily on the technical attributes of phishing attacks, such as the content and appearance of spoofed website URLs or of phishing emails and text. In contrast, individual aspects of users and risk communication/mental models were rarely explored in-depth (five out of 51 papers). The majority of research (32 out of 51 papers) sought to develop security indicators or warning tools that were beneficial for the users; however, in these studies, important details about the participants were invariably missing. Demographic and personal factors such as technical expertise may play a vital role in individual reactions to security threats (Das et al., 2018). Thus, these details are important for any conclusive user-focused studies. Yet even the researchers in our data set who developed training tools or games which focused on risk communication seldom discussed the details of their users and any behavioral variations that might be related to these.

Our analysis and findings reveal that the majority of the researchers noted that URLs and the appearance of spoofed websites were a key indicator of phishing attempts. Thus, they proposed new security tools, such as browser extensions and warning indicators, etc., to reduce the likelihood of users falling victim to phishing schemes. While it is extremely important to provide technical solutions, we believe that understanding the human factors that allow a phisher to successfully exploit the user is crucial for detection, prevention, and mitigation strategies. Risk communication is an emerging field in phishing that seeks to employ more efficient training methods for just these reasons. Four of the papers studied provided game-based risk communication training as a solution to mitigate such threats, for instance. However, to develop the most effective collection of such tools and strategies, more research should focus on understanding the mental models, situational factors, and related behaviors of users, a strategy embraced by only one study in our data set.

In order to offset the deficiencies, we found both in the reporting and, most likely, the design of phishing research, our literature review suggests a two-fold plan of action for future user-based studies in phishing. First, researchers should report accurate demographic and other study-relevant information on their participants. Second, assuming that the current lack of reporting reflects a lack of intentional, strategic recruitment strategy, future researchers should make every effort to recruit more diverse and representative samples of participants. In order to generate useful research, the main goal of the recruitment phase of a study should be to recruit a participant pool that will mirror the audience who is intended to benefit from the phishing research goal. This may include a commitment to a participant group with, at a minimum, an equal gender balance, a wide range of ages and educational attainment, varying racial, ethnic, and cultural backgrounds, and multiple levels of technical literacy. Only then can researchers begin to pay careful attention to any systematic variation between these attributes and participants' reported behaviors, goals, and mindsets. These adjustments to future work will create more effective phishing research, as they will begin to encompass the wide range of human demographics that phishers attempt to exploit.

## 6. Conclusion

Phishing attacks are one of the oldest known cyber-attacks, resulting in the loss of billions of dollars every year (Moore and Clayton, 2011; Tian et al., 2018). In 2018, the Federal Bureau of Investigation estimated that companies around the world lost $12 billion because of business email compromises (Digitalinformationworld.com, 2019). Preventing phishing attacks is a high priority and a major challenge in the domain of secure computing. While researchers and practitioners often provide technical solutions to try to solve phishing-related issues, our analysis of phishing research suggests that social solutions, focused on users, themselves, might offer an important, missing piece of the complete toolkit available to counter these attacks. Phishing research papers from the ACM Digital Library revealed that only 13.9% of relevant published papers from 2004 to 2018 included any kind of user-focused study, and these studies primarily focused on usability or testing of tools developed by the researchers rather than exploring the ways different kinds of users approach and make sense of phishing attempts. Much of this research also failed to include crucial details about study participants. Based on our analysis of published phishing research to date, we find support for the potential importance of user studies in this field of research and for, better reporting and recruiting practices in future studies within this field.

## Acknowledgments

We would like to thank Gabriel Lahman, Stephen Railing, and Ke Xu for vital assistance in the initial data collection and creation of the codebook. We are grateful to Dr. L. Jean Camp for insightful guidance on this paper, and to Dr. Sameer Patel for early conversations about related work. Any opinions, findings, and conclusions or recommendations expressed in this material are those of the author(s) and do not necessarily reflect the views of the Indiana University Bloomington.